\documentclass[pra,twocolumn,aps,showpacs]{revtex4-1}

\usepackage{amsmath}
\usepackage{amssymb}
\usepackage{graphicx}
\usepackage{color}
\usepackage{bm}
\usepackage{times}
\usepackage{hyperref}
\hypersetup{
  colorlinks=true,
  citecolor=blue,
  linkcolor=blue,
  urlcolor=blue}

\definecolor{purp}{RGB}{228, 35, 180}
\begin{document}


\title{Ramifications of topology and thermal fluctuations in quasi-2D 
       condensates}

\author{Arko Roy}
\author{D. Angom}
\affiliation{Physical Research Laboratory,
             Navrangpura, Ahmedabad-380009, Gujarat,
             India}

\begin{abstract}
We explore the topological transformation of quasi-2D Bose-Einstein 
condensates of dilute atomic gases, and changes in the low-energy 
quasiparticles associated with the geometry of the confining potential. 
In particular, we show the density profile of the condensate and quantum 
fluctuation follow the transition from a multiply to a simply connected 
geometry of the confining potential. The thermal fluctuations, in contrast,
remain multiply connected. The genesis of the key difference lies in the
structure of the low-energy quasiparticles. For which we use the 
Hartree-Fock-Bogoliubov, and study the evolution of quasiparticles, 
the dipole or the Kohn mode in particular. We, then employ the 
Hartree-Fock-Bogoliubov theory with the Popov approximation
to investigate the density and the momentum distribution of the thermal
atoms. 
\end{abstract}


\maketitle

\section{Introduction}
  Toroidal or multiply connected condensates are splendid model systems to
study Kibble-Zurek mechanism~\cite{kibble_76,zurek_85,zurek_96} in detail. 
Understanding of 
this mechanism in Bose-Einstein condensates (BECs), which explains
spontaneous seeding of topological defects during phase-transitions, has 
attracted much attention. There has been several, both theoretical 
and experimental, recent works on this topic~\cite{weiler_08,das_12,
lamporesi_13,su_13}. 
In this context, an investigation on the nature of quantum and thermal
fluctuations in such systems is of importance, and can provide better insights 
to the physics of defect formation. Furthermore, toroidal BECs are
significant as waveguides in atom interferometers~\cite{marti_15}, 
which are analogs of SQUIDs~\cite{ryu_13,mathew_15}.

  The remarkable experimental realizations of toroidal condensates through 
a variety of sophisticated and novel techniques have opened up new 
possibilities to explore multiply connected BECs with finer control, and in
better detail than ever before. To mention a few, toroidal condensates 
have been obtained with the use of harmonic potential in combination with
a Gaussian potential~\cite{ryu_07}, Laguerre-Gaussian 
beams~\cite{ramanathan_11,moulder_12,curtis_03,beattie_13}, combination of
an RF-dressed magnetic trap with an optical 
potential~\cite{heathcote_08,morizot_06},
magnetic ring traps~\cite{sauer_01,gupta_05,arnold_06,sherlock_11},
time-averaged ring potentials~\cite{henderson_09,bell_16}, coincident red
and blue detuned laser beams~\cite{marti_15}, and employing digital micromirror
devices ~\cite{kumar_16}. For the present work, the realization consisting
of a harmonic and Gaussian potential~\cite{ryu_07} is of importance as it 
offers the possibility of transforming harmonic (simply connected) to 
toroidal (multiply connected) confining potential by modifying the Gaussian 
potential. This is, in other words, equivalent to a pancake shaped BEC
getting transformed to a toroidal one. The other, equally important, topics
are the effects associated with the transition from multiply to simply 
connected BEC due to relative shift in the constituent trapping potentials.
Here, the question ``How do the fluctuations change as a multiply connected
BEC is transformed to a simply connected one?" is a pertinent one. 
The answer to this question has profound experimental implications since, 
in experiments the trap centers never coincide. This is because of 
gravitational sagging, and deviations of the trapping potentials from 
perfect alignment.

  In the present work, we use the Hartree-Fock-Bogoliubov theory with
Popov (HFB-Popov) approximation to gain insights on the quantum and thermal 
fluctuations as
the separation between the trap centers increases. Starting from a
perfectly aligned trapping potential to a misaligned configuration, our
studies reveal that the multiply connected BEC gets transformed to a 
simply connected one. This is accompanied with the breaking of rotational
symmetry of the system, and hardening of the quasiparticle excitations.
Furthermore, we show that the quantum fluctuations have the same geometry
as the BEC when the transformation from multiply to simply 
connected geometry occurs. But, the thermal fluctuations retains
the multiply connected geometry. These are reflected in the momentum
distribution of the quantum and thermal non-condensate densities. This 
indicates that the spontaneous seeding of topological defects can have
different distributions depending on the geometry of the confining potential.

The paper is organized as follows: In Sec.~\ref{theory}, we briefly explain the 
HFB-Popov formalism for interacting quasi-2D BEC and provide a description
of how to compute the effective radial trapping frequencies associated with the
change in the geometry of the confining potential. 
The results and discussions are given in Sec.~\ref{results}.
The evolution of the quasiparticle excitation energy and amplitudes 
corresponding 
to the Kohn mode are presented in Sec.~\ref{kmode}. The transition from 
multiply to simply connected geometry of the trapping potential brings about 
a change in the nature of quantum and thermal fluctuations and momentum 
distribution of the thermal atoms which are illustrated 
in Sec.~\ref{qtf},~\ref{md}. The dispersion curves are then presented 
in Sec.~\ref{dsp}. We, then, end with conclusions highlighting the key 
findings of the present work in Sec.~\ref{conc}.

\section{HFB-Popov approximation in quasi-2D BEC}
\label{theory}
The second quantized form of the grand-canonical Hamiltonian describing
an interacting quasi-2D BEC confined with trapping potential $V(x,y)$ is 
\begin{eqnarray}
  \hat{H} &=& \iint dx dy\,\hat{\Psi}^\dagger(x,y,t)
        \bigg[-\frac{\hbar^{2}}{2m}\left(\frac{\partial ^2}{\partial x^2} +
        \frac{\partial ^2}{\partial y^2}\right)
        + V(x,y)\nonumber\\
      &&  -\mu + \frac{U}{2}\hat{\Psi}^\dagger(x,y,t)\hat{\Psi}
        (x,y,t)\bigg]\hat{\Psi}(x,y,t),
\label{hamiltonian} 
\end{eqnarray}
where, $\hat{\Psi}$ and $\mu$ are the Bose field operator of the single species 
BEC, and the chemical potential, respectively. Starting from a general 3D 
harmonic confining potential $V(x,y,z)=(1/2)m\omega_x^2(x^2 + \alpha^2y^2 +
\lambda^2z^2)$, we obtain a rotationally symmetric quasi-2D system when
the anisotropy parameters are $\alpha = \omega_y/\omega_x=1$ and 
$\lambda = \omega_z/\omega_x \gg 1$. The excitations along $z$ are then
suppressed, and along this axis, the condensate remains in the ground state. 
Hence, we can integrate out the $z$ dependence, and the confining potential
is reduced to $V(x,y)=(1/2)m\omega_\perp^2(x^2 + y^2) $ and obtain the above
Hamiltonian. With these considerations, the excitations and dynamics are 
limited to the $xy$-plane. Furthermore, the atoms repulsively interact with 
strength $U = 2 a\sqrt{2\pi\lambda}$, with $a$ as the $s$-wave scattering 
length, and $m$ as the atomic mass. From the Hamiltonian, using the 
variational method with Bogoliubov approximation, we obtain a pair of
coupled equations. These are the generalized Gross-Pitaevskii (GP) equation, 
and Bogoliubov-de Gennes equations. Together, the equations describe the 
equilibrium state of the condensate and non-condensed cloud of atoms in
the confining potential at finite temperatures. For the present work, we
use Hartree-Fock-Bogoliubov (HFB) theory to calculate the condensate
and non-condensate density distributions. Further more, we use Popov
approximation (HFB-Popov) to obtain gapless excitation spectra
~\cite{gies_04,gies-04,roy_14,roy_14a,roy_15a}, and hence, 
maintain the Hugenholtz-Pines theorem~\cite{hugenholtz_59}. 

In the HFB-Popov approach, the Bose field operator $\hat\Psi$ is a linear
combination of the $c$-field or the condensate part represented by 
$\phi(x, y, t)$, and the non-condensate or the fluctuation part 
denoted by $\tilde\psi(x,y,t)$~\cite{griffin_96}. 
That is, $\hat{\Psi} = \phi + \tilde\psi$
and the equation of motion of $\phi(x, y, t)$, the generalized GP equation, is 
\begin{equation}
  \hat{h}\phi + U\left[n_{c}+2\tilde{n}\right]\phi = 0,
  \label{gpe1s}
\end{equation}
where, $\hat{h}= (-\hbar^{2}/2m)\left(\partial ^2/\partial x^2
                 +\partial ^2/\partial y^2\right) + V(x,y)-\mu$ represents 
the single-particle or the non-interacting part of the Hamiltonian.
We obtain the generalized GP equation by reducing the three-field correlation 
term into a quadratic form in fluctuation operators based on Wick's theorem, 
and then, taking average to obtain the stationary state 
solution~\cite{griffin_96}. We define $n_{c}(x,y)\equiv|\phi(x,y)|^2$,
$\tilde{n}(x,y)\equiv\langle\tilde{\psi}^{\dagger}(x,y,t)
\tilde{\psi}(x,y,t)\rangle$, and $n(x,y) = n_{c}(x,y)+ \tilde{n}(x,y)$
to be the local condensate, non-condensate, and total density,
respectively. We use Bogoliubov transformation such that the fluctuations 
operator, in terms of the quasiparticle modes, are the following 
\begin{eqnarray}
   \tilde{\psi} &=&\sum_{j}\left[u_{j}(x,y)
    \hat{\alpha}_j(x,y) e^{-iE_{j}t/\hbar}
   - v_{j}^{*}(x,y)\hat{\alpha}_j^\dagger(x,y) e^{iE_{j}t/\hbar}
    \right],\nonumber\\
\label{ansatz}
\end{eqnarray}
with $\hat{\alpha}_j$ ($\hat{\alpha}_j^\dagger$) as the quasiparticle
annihilation (creation) operators which satisfy the usual Bose commutation
relations, and the subscript $j$ denotes the mode index. Here, $u_j$ and 
$v_j$ are the Bogoliubov quasiparticle amplitudes of the
$j$th mode. From these definitions and considerations we arrive at the 
following pair of coupled Bogoliubov-de Gennes (BdG) equations
\begin{subequations}
\begin{eqnarray}
(\hat{h}+2Un)u_{j}-U\phi^{2}v_{j}&=&E_{j}u_{j},\\
-(\hat{h}+2Un)v_{j}+U\phi^{*2}u_{j}&=& E_{j}v_{j}.
\end{eqnarray}
\label{bdg1}
\end{subequations}
We obtain the above equations by expressing the equation of motion of 
$\tilde{\psi}$ as that of $\hat{\Psi}$ with the subtraction of $\phi$. To 
solve, the two equations are treated as matrix equation with eigenstates of
the harmonic oscillator potential as the basis. The eigenvalues and 
eigenvectors of the matrix, obtained from diagonalizing the matrix,  are then 
the quasiparticles or mode energies and amplitudes. 

 Once the quasiparticle energies and amplitudes are known, the thermal or 
non-condensate density $\tilde{n}$ at temperature $T$ is 
\begin{equation}
 \tilde{n}=\sum_{j}\{[|u_{j}|^2+|v_{j}|^2]N_{0}(E_j)+|v_{j}|^2\},
  \label{n_tilde}
\end{equation}
where $\langle\hat{\alpha}_{j}^\dagger\hat{\alpha}_{j}\rangle = (e^{\beta
E_{j}}-1)^{-1}\equiv N_{0}(E_j)$ with $\beta=1/k_{\rm B} T$, is the Bose
factor of the $j$th quasiparticle state with energy $E_j$ at temperature $T$.
For $T\rightarrow0$, the non-condensate density arises out of the quantum
fluctuations when $ N_0(E_j)$'s in Eq. (\ref{n_tilde}) vanish. The 
non-condensate density is then reduced to $ \tilde{n} = \sum_{j}|v_{j}|^2$.
The essence of HFB-Popov theory is to obtain self-consistent solutions of 
the coupled equations Eq. (\ref{gpe1s}) and Eq. (\ref{bdg1}).

\begin{figure}[h]
 \includegraphics[width=8.5cm]{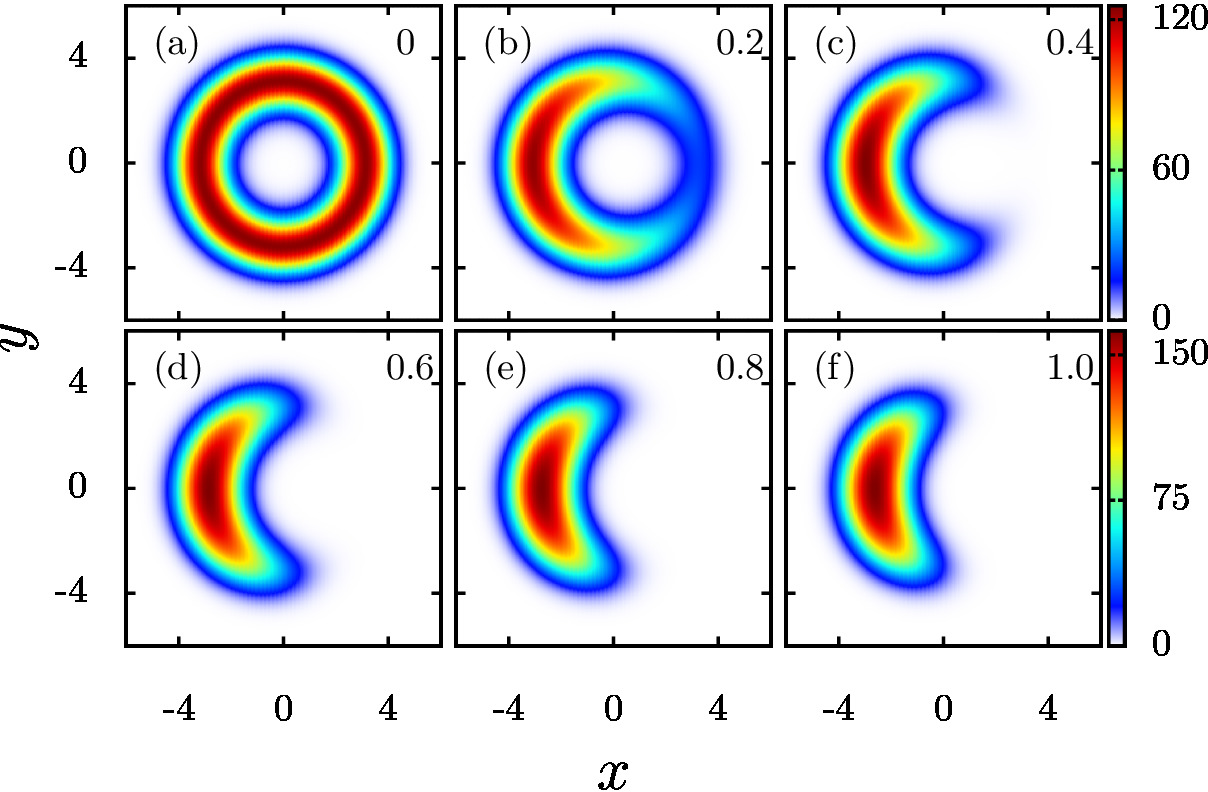}
   \caption {(Color online) Plots of $n_c$ at $T=0$ for 
             different
             values of $\Delta_x$ (shown at top right corner of each image). 
             Shows the transformation of $n_c$ from toroidal to bow-shaped
             structure. (a) For $\Delta_x=0$, $n_c$ assumes 
             the form of a toroid which has a multiply-connected geometrical 
             structure. (b) With $\Delta_x=0.2$, the rotational symmetry of the
             condensate is broken, and (c) - (f) shows the simply-connected 
             profiles of $n_c$ when the trap centers are 
             non-coincident. Here $x$, $y$, and $\Delta_x$ are measured 
             in units of $a_{\rm osc}$. In the plots $n_c$  
             are in units of $a_{\rm osc}^{-2}$.
            }
\label{den}
\end{figure}


\subsection{Transition from Harmonic to toroidal potential}

 In the experiments with multiply connected BECs, toroidal in the present 
work, the confining potential could be a superposition of harmonic and Gaussian 
potentials with a common center~\cite{ryu_07}. However, in practice, the 
trap centers do to coincide due to gravitational sagging or due to the fact 
that perfect alignment of optical and mechanical elements in experiments is 
improbable. Taking this into account, the net confining potential is 
\begin{equation}
V_{\rm net}(x,y) = V(x,y) + U_0e^{{-[(x-\Delta_x)^2 + \alpha^2 y^2}]/2\sigma^2}
\end{equation}
where, $V(x,y)$ is the harmonic potential described earlier, and second term
is the Gaussian potential. In the latter, $U_0$ and $\sigma$ are the 
amplitude  and width, respectively. To simplify, we consider there is a 
relative separation $\Delta_x$ along $x$-axis between the minima of the 
harmonic 
potential $V(x,y)$ and maxima of the 2D-Gaussian potential. For the case of
$U_0=0$ and $\Delta_x=0$, when $\alpha=1$, $V_{\rm net}$ is rotationally 
symmetric. For $U_0\gg0$ and $\Delta_x=0$, $V_{\rm net}$ deviates from harmonic
potential, and at larger values of $U_0$ the potential assumes the form of 
a doughnut or a toroid (multiply connected). The rotational symmetry is, 
however, broken for $\Delta_x\neq0$. An intriguing possibility is to increase 
$\Delta_x$, such that the two minima along the $x$ axis are well separated in 
energies. As the $\Delta_x$ is increased, the geometry of the BEC is then 
transformed from toroidal to a bow-shaped structure as shown in 
Fig. \ref{den}. In the present work,
we address the issues of how the relative shift affects the quasiparticles 
and the non-condensate densities. It must be mentioned here that, small
values of $\Delta_x$ are important for realistic theoretical studies on 
finite temperature effects, and a continuous variation to large $\Delta_x$ 
offers 
a possibility to examine dynamics of topological defects, persistent 
currents, interplay of coupling between condensate and thermal clouds, and 
most importantly, evolution of mode energies and amplitudes. For the latter,
the low energy excitations, the Kohn mode in particular, are of especial 
importance as quantum and thermal fluctuations have predominant contributions 
from these modes.

   Although, we have considered toroidal potential as superposition of
harmonic and Gaussian potentials, the other possibility is using 
Laguerre-Gaussian (${\rm LG}_{p}^{l}$) laser beams. These were first
examined in theoretical works \cite{wright_2000,courtial_97},
and eventually toroidal condensates of atomic $^{23}$Na
\cite{ramanathan_11}, and $^{87}$Rb~\cite{moulder_12} have been experimentally
achieved using ${\rm LG}_0^l$ beams~\cite{curtis_03}.
It is worth mentioning here that in our earlier work \cite{roy_15b}, 
we demonstrated 
an alternative scheme of obtaining multiply connected BECs and its effects on 
the fluctuations. In the calculations, the spatial and temporal variables are 
scaled as $x/a_{\rm osc}$, $y/a_{\rm osc}$ and $\omega_x t$ respectively, 
where $a_{\rm osc} = \sqrt{\hbar/m\omega_x}$.


\subsection{Effective radial trapping frequencies}

 One parameter which provides an insight on the condensate density
distribution $n_c$ with the variation in $\Delta_x$ is the effective radial 
trapping 
frequency $\omega_r$ of $V_{\rm net}$. Consider the case of $\Delta_x=0$, 
the $V_{\rm net}$ is then a rotationally symmetric Mexican hat potential, and 
$\omega_r$ about the minima of $V_{\rm net}$ is independent of the azimuthal 
angle $\phi$. However,  $V_{\rm net}$ breaks rotational symmetry when 
$\Delta_x \neq 0$ and the general form of the effective radial frequency is 
$\omega_r(\phi)$, it is a function of $\phi$. For $\Delta_x>0$, with the 
relative 
shift along the $x$-axis the the highest and lowest effective radial 
frequencies are $\omega_r(0)$ and $\omega_r(\pi)$, respectively.  
To compute these frequencies, let $x_0$ and $x_\pi$ denote the two minima 
of $V_{\rm net}$ along the $x$-axis. This is a (a)symmetric double-well 
potential when ($\Delta_x>0$) $\Delta_x=0$, and with $\delta=x-x_{0,\pi}$, 
a Taylor series expansion about $x_{0,\pi}$ gives the effective harmonic
oscillator potential around it as
\begin{equation}
  V_{\rm eff}(\delta) = \left\{1 + \frac{U_0}{\sigma^2}
  e^{-(x_{0,\pi}-\Delta_x)^2/2\sigma^2} \left[\frac{(x_{0,\pi} -
  \Delta_x)^2}{\sigma^2} - 1\right]\right\} \delta ^2,
\end{equation}
and the effective radial frequencies are 
$\omega_r^2(0)= 1 + \frac{U_0}{\sigma^2}
  e^{-(x_{0}-\Delta_x)^2/2\sigma^2} \left[\frac{(x_{0} -
    \Delta_x)^2}{\sigma^2} - 1\right] $, and
$\omega_r^2(\pi)=1 + \frac{U_0}{\sigma^2}
  e^{-(x_{\pi}-\Delta_x)^2/2\sigma^2} \left[\frac{(x_{\pi} -
    \Delta_x)^2}{\sigma^2} - 1\right]$. The 
geometry of the condensate, and quasiparticle spectra depends on the relative
values of these two frequencies. In addition, another important parameter
is the energy difference 
$\Delta E = V_{\rm net}(x_0,0) -  V_{\rm net}(x_\pi,0)$ between the minima
along $x$-axis. Based on Thomas-Fermi approximation, where $n_c$ 
is proportional to the confining potential, the condensate density
profile is simply connected when $\mu < \Delta E$.
\begin{figure}[h]
 \includegraphics[width=8.5cm]{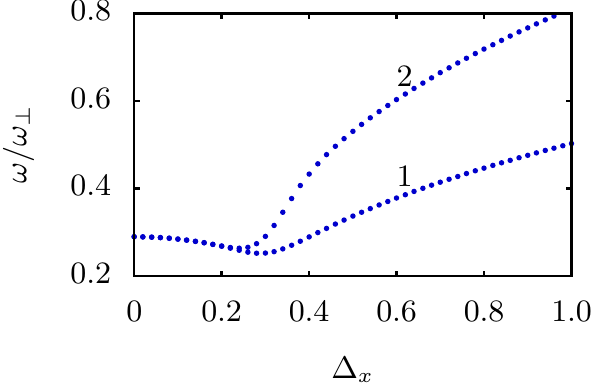}
 \caption {(Color online) Evolution of Kohn mode energies as a function
           of $\Delta_x$ at $T=0$. Shows the lifting of degeneracy of Kohn mode
           energy in the domain $0\leqslant \Delta_x \leqslant 1$ as a result of
           topological deformation. The quasiparticle
           amplitudes corresponding to the energy branches `1' and `2'
           are structurally different. Here $\Delta_x$ is measured in units
           of $a_{\rm osc}$. 
          }
\label{mode_evol_sol}
\end{figure}


\subsection{Dispersion Relations}
The dispersion relation of a physical system determines how it responds to 
external perturbations, in particular, those which can generate density 
variations at scales much smaller than the system size. For the present work,
the transformation in the geometry of the external trapping potential from 
harmonic to toroidal, and the consequent transition of the BEC from multiply 
to simply connected topology affects the energy of the quasiparticle 
excitation.  Hence, the dispersion relation is expected to change. For this, 
as the BEC is of finite size we compute the root mean square of the
wavenumber $k^{\rm rms}$ of each quasiparticle mode to define a discrete
dispersion relation. Following Ref.~\cite{wilson_10}, the 
$k^{\rm rms}$ of the $j$th quasiparticle is
\begin{equation}
 k_j^{\rm rms} = \left\{\frac{\int d\mathbf{k} k^2
                 [|u_j(\mathbf{k})|^2 + |v_j(\mathbf{k})|^2]}
                 {\int d\mathbf{k} [|u_j(\mathbf{k})|^2 
                 + |v_j(\mathbf{k})|^2]}\right\}^{1/2}.
  \label{dspeq}
\end{equation}
Here, it is to be noted that $k_j^{\rm rms}$ are in terms of the 
quasiparticle modes defined in the momentum space, and hence, it is essential
to compute $u_j(\mathbf{k})$ and $v_j(\mathbf{k})$, the Fourier transform of 
the Bogoliubov quasiparticle amplitudes $u_j(x,y)$ and $v_j(x,y)$, 
respectively. Once we have $k_j^{\rm rms}$ for all 
the modes we obtain a dispersion curve, and we can then examine how the 
change in the condensate topology modifies the dispersion curve. In earlier
works, dispersion curves have been obtained and examined for
harmonically trapped binary BECs ~\cite{ticknor_14},
binary BECs in optical lattices ~\cite{suthar_16}, and 
dipolar BECs~\cite{blakie_13,bisset_13}. In this work we address how the 
topology of the BEC modifies the dispersion relation.
\begin{figure}[t]
 \includegraphics[width=8.0cm]{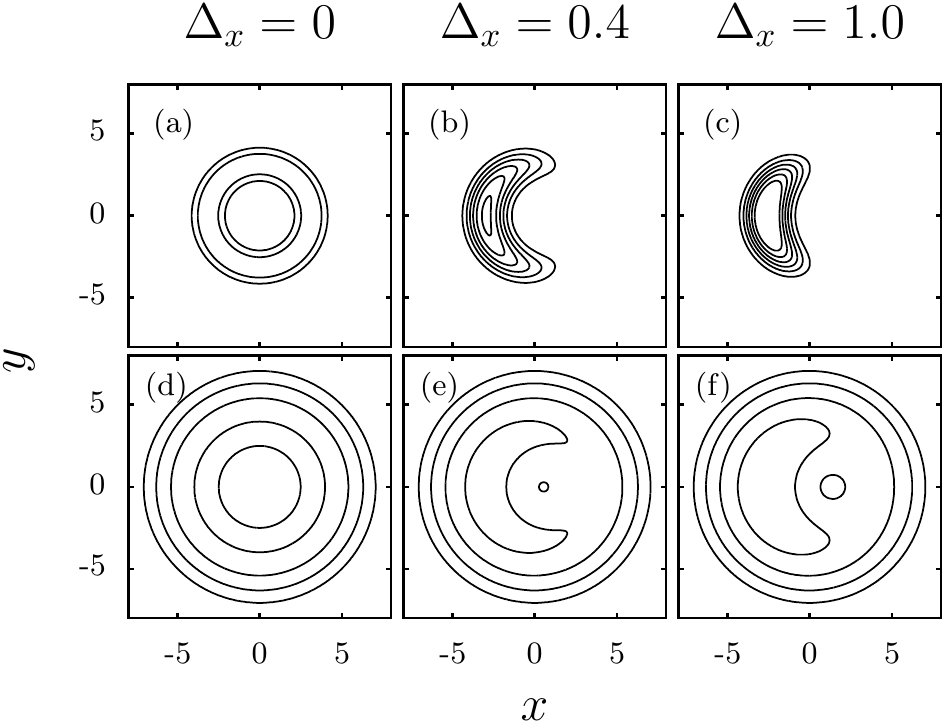}
 \caption {(Color online)(a)-(c) The upper panel shows the contours of
            $n_c$, and (d)-(f) the lower panel shows the equipotential 
            curves corresponding to three different values of $\Delta_x$.
            Here $x$, $y$ and $\Delta_x$ are measured in units
            of $a_{\rm osc}$.
          }
\label{contour}
\end{figure}


\section{Results and Discussions}
\label{results}
 To probe the effects of topological modification of the quasi-2D BEC from 
multiply to simply connected geometry, we solve the equations in HFB-Popov 
theory numerically. The results discussed are generic to BEC of any atoms 
with slight variations in the parameters, and for a detailed examination
we consider the specific case of $^{23}$Na BEC with $a =
53.3a_0$. Experimentally, toroidal condensates of  $^{23}$Na atoms have
been achieved by combining a harmonic potential with a Gaussian
potential~\cite{ryu_07}. 
The evolution of the quasiparticle modes is computed for 
$N_{\rm Na} = 2\times 10^3$ with
$\lambda = 39.5$, $\omega_x=\omega_y=\omega_{\perp} = 2\pi \times 20.0$ Hz, and 
$U_0 = 15\hbar\omega_{\perp}$. At the outset when $U_0 =
15\hbar\omega_{\perp}$ and $\Delta_x=0$, the density
profile has rotational symmetry with $n_c(0,0)\approx0$ as shown in
Fig.~\ref{den}(a). The condensate cloud assumes the form of a toroid.
The low-lying excitation spectrum is characterized by the presence of 
doubly degenerate $m=1$ modes, and among these, the most important ones
are the Kohn modes with $\omega/\omega_\perp=0.29$. 

  For studying mode evolutions, we do a series of computations using HFB-Popov
approximation with increasing $\Delta_x$ to obtain the quasiparticle spectra 
and fluctuations. When $\Delta_x\neq 0$ the rotational symmetry of the 
potential and hence, the condensate cloud are broken. As $\Delta_x$ is 
increased, the profile of $n_c$ is transformed from toroidal to bow-shaped 
geometry. This corresponds to a transition from multiply to simply connected
topology, and the transition is shown in Fig.~\ref{den} with a set of 
selected values of $\Delta_x$. The key point to note here
is the modification of the excitation spectra and the structure of the
Bogoliubov quasiparticle amplitudes. One of the most distinctive features 
is the evolution of the low-energy $m=1$ modes. From these modes we select
the one with lowest energy, the Kohn mode, and the evolution is  as shown 
in Fig.~\ref{mode_evol_sol} using HFB theory. 
\begin{figure}[t]
 \includegraphics[width=8.5cm]{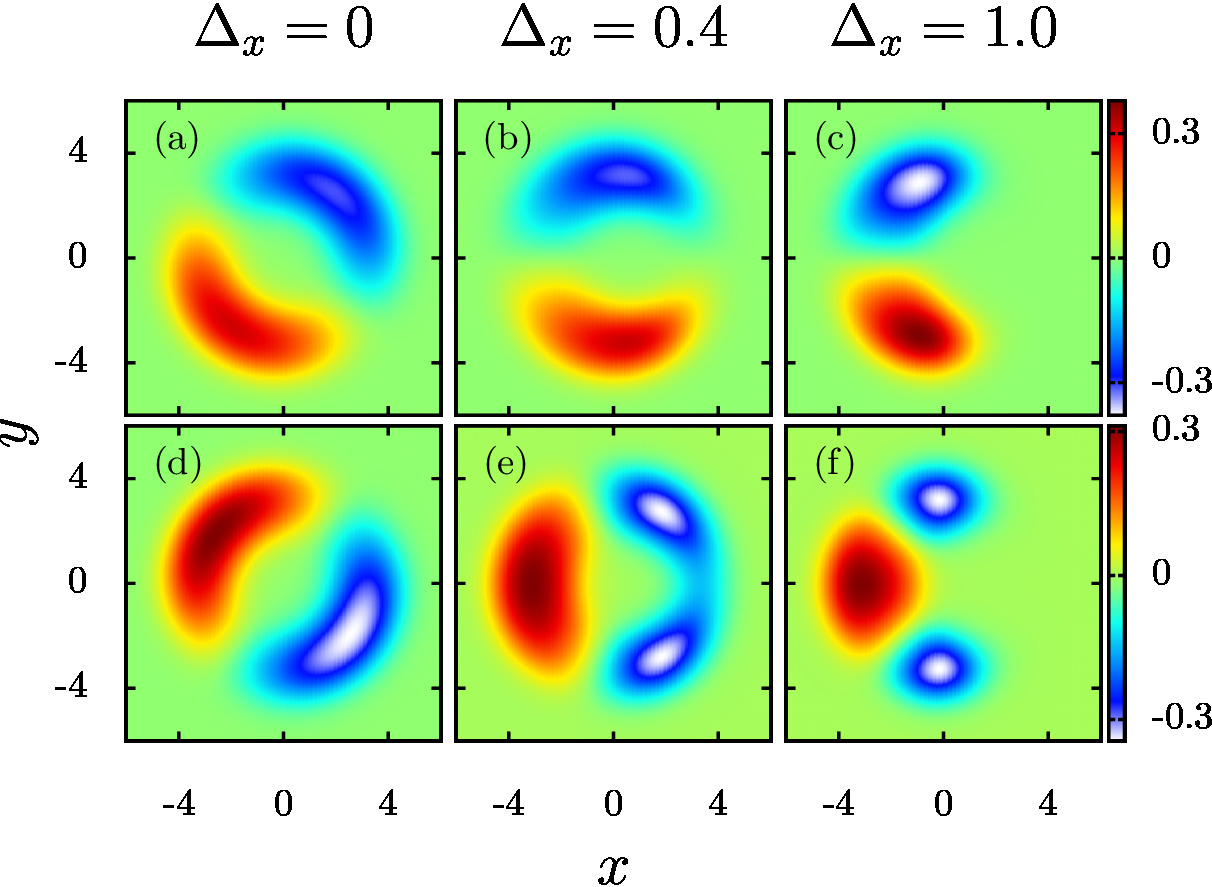}
 \caption {(Color online) Evolution of quasiparticle amplitudes
           corresponding to the dipole or Kohn mode as $\Delta_x$ is varied 
           from 0 to $1{\rm a_{osc}}$. 
           (a) - (c) Show the $u_{\rm Na}$ corresponding
           to the Kohn mode energies identified as `$1$' in
           Fig.~\ref{mode_evol_sol}. (d) - (f) Show the $u_{\rm
           Na}$ corresponding to the Kohn mode energies identified as `$2$' 
           in Fig.~\ref{mode_evol_sol}. In the plots $u$ and $v$ are in 
           units of $a_{\rm osc}^{-1}$. Here $x$, $y$ and $\Delta_x$ are 
           measured in units of $a_{\rm osc}$.
          }
\label{k12}
\end{figure}


\subsection{Kohn mode evolution} 
\label{kmode}
 The Kohn modes, among the low-energy excitations, is one of the most 
important as it has leading contribution to quantum and thermal fluctuations.
For this reason and as a case study, we examine it's evolution in detail, 
however, the trends observed apply to other low-lying $m=1$ modes as well.  
The Kohn mode is doubly degenerate when $\Delta_x=0$, and as shown in 
Fig. \ref{mode_evol_sol}, the degeneracy continues with decreasing energy in 
the domain $0<\Delta_x\lessapprox 0.22$. In this domain, the geometry of the 
condensate is multiply connected, but the density maxima is shifted radially 
outward. Thus, it increases the wavelength of the excitations which lie along 
the toroidal axis, and explains the decrease in energy of the Kohn mode, 
and the other low-lying modes with $m=1$. These observations are, as mentioned 
earlier, coupled to the form of the confining potential, and evident from the 
equipotential curves of $V_{\rm net}$ and density contours shown 
in Fig.~\ref{contour}(a) and (b).

 At a critical value of $\Delta_x$ the toroidal condensate is transformed to 
a simply connected geometry. As a result, for $\Delta_x > 0.22$ the degeneracy
of the Kohn modes is lifted, and the mode energy bifurcates into two 
branches marked as `1' and `2' in Fig.~\ref{mode_evol_sol}. In addition, the 
two modes harden, and there are discernible changes in the structure of
the mode functions. As shown in Fig.~\ref{k12}(a), at $\Delta_x=0$, the 
Kohn mode functions are $\pi/2$ rotation of each other and mutually orthogonal.
However, with increase in $\Delta_x$ and after bifurcation, the mode function
of the lower energy or the branch marked `1' in Fig.~\ref{mode_evol_sol}
retains the dipole structure Fig.~\ref{k12}(a)-(c). But, the wave number
increases, and hence, the two lobes gets smaller. For the other mode 
function with energy marked as `2' in Fig.~\ref{mode_evol_sol}, one of the 
lobes begins to cleave for $\Delta_x > 0.22$, and as shown in 
Fig.~\ref{k12}(e) there are three distinct lobes at $\Delta_x \approx 0.4$. At
higher values, say $\Delta_x=1.0$, the three lobes are well separated, and 
it is effectively transformed into a mode which is neither $m=1$
or $m=2$, and its $m$ may be considered as hybridization of $1$ and $2$. 
Thus, the increase in wave number, and transformation to a mode whose $m$
is greater than unity accounts for the increase in the energies of the 
two modes. 
\begin{figure}[t]
 \includegraphics[width=8.5cm]{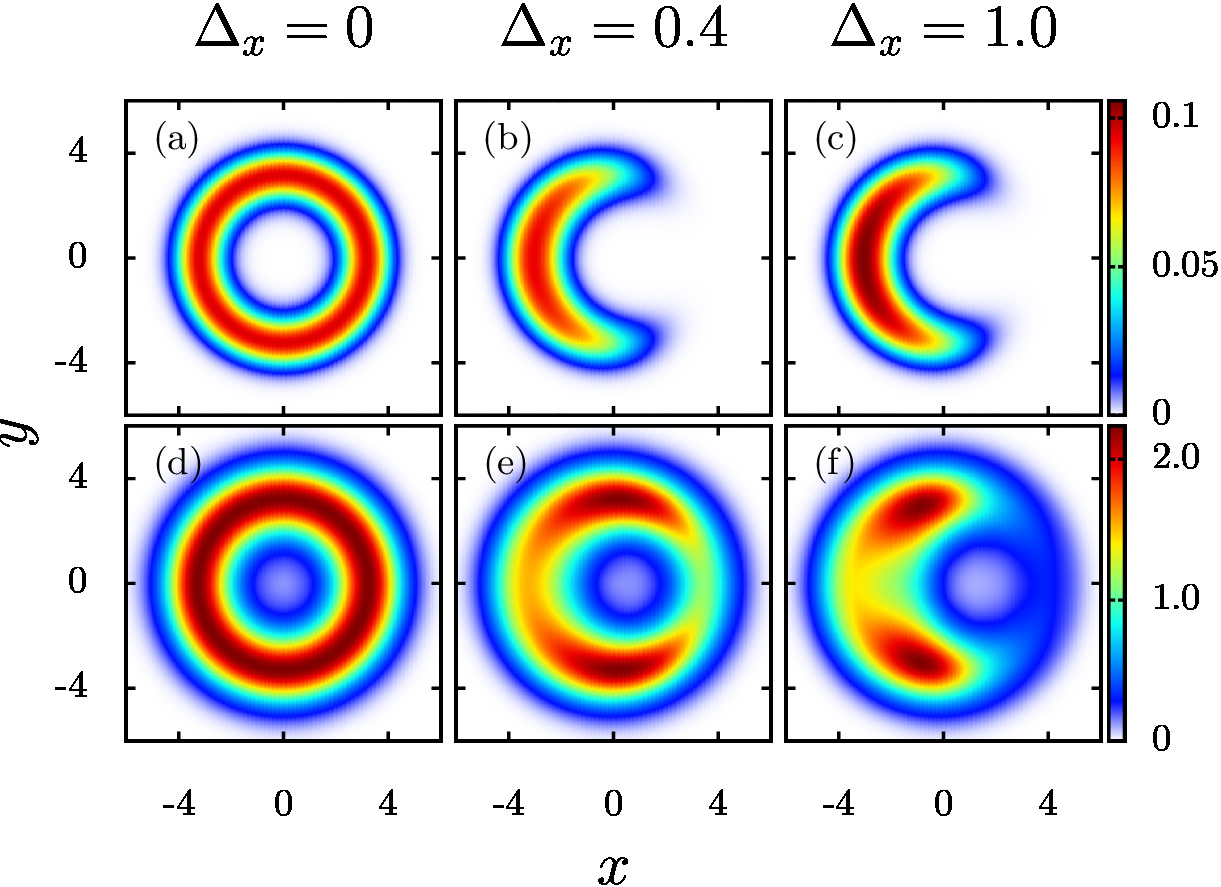}
  \caption {(Color online) Condensate and non-condensate density profiles 
            for different values of $\Delta_x$. 
            (a) - (c) plots of non-condensate density due to 
            quantum fluctuations at $T=0$, and (d) - (f) thermal density
            distribution at $T=10$ nK for $\Delta_x = 0, 0.4, 1.0 a_{\rm osc}$,
            respectively. Here $x$, $y$, and $\Delta_x$ are measured in units
            of $a_{\rm osc}$. In the plots $\tilde{n}$ 
            are in units of $a_{\rm osc}^{-2}$.
            }
  \label{qt}
\end{figure}


\subsection{Quantum and thermal fluctuations}
\label{qtf}
 Quantum fluctuations or zero temperature fluctuations, are intrinsic to any 
interacting quantum many-body system, and in the HFB-Popov approximation 
$|v|^2$ of the Bogoliubov quasiparticle amplitudes contribute to 
the quantum fluctuations. We find that with larger $\Delta_x$, 
as the multiply connected condensate attains a simply connected geometry, 
there is a reduction in the total number of non-condensate atoms 
$\tilde{N} = \int \tilde{n}(x,y)\,dxdy$ with $\tilde{n}(x,y)$ as the 
non-condensate atom density. It is to be emphasized that $\tilde{n}(x,y)$ 
arising from quantum fluctuations coincides with the $n_c$
for  $\Delta_x\geqslant 0$, and specific examples are displayed in 
Figs.~\ref{qt} (a)-(c).

  For $T\neq0$, both the quantum and thermal fluctuations contribute to 
$\tilde{n}$, however, the latter has a much larger contribution.
When $\Delta_x=0$, the condensate and the thermal clouds have similar
density profiles, and possess overlapping maxima as shown in Fig.~\ref{qt} (d). 
But, when $\Delta_x > 0$ the profiles evolve very differently with
increase in $\Delta_x$; the profile of $n_c$ is transformed to
simply connected geometry, whereas $\tilde{n}$ retains
multiply connected structure as shown in Figs.~\ref{qt} (e)-(f). In other
words, there is a striking difference in the thermal component of
$\tilde{n}$ to that of the quantum fluctuations. This key difference is
due to the thermal atoms filling up the region with depleted condensate 
density, and arises from the contributions to thermal fluctuations from 
quasiparticles with energies $E_j > \Delta E$. Where, $\Delta E$ as defined
earlier is the energy difference between the two minima along the 
$x$-axis $V_{\rm net}(x_0,0) -V_{\rm net}(x_\pi,0)$.
\begin{figure}[t]
 \includegraphics[width=8.5cm]{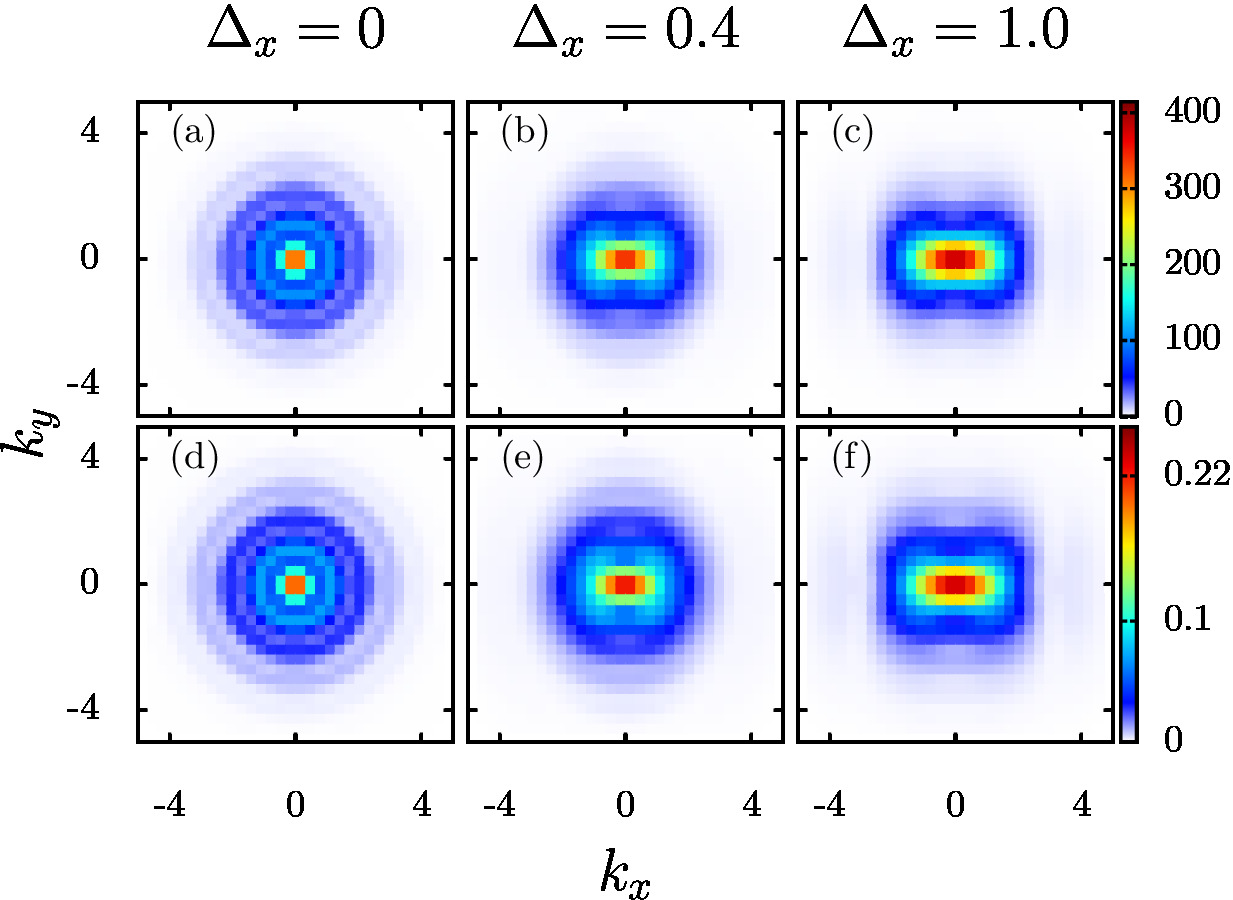}
 \caption {Momentum distribution of the densities for different values
           of $\Delta_x$ at $T=0$. (a)-(c) Show the momentum distribution 
           corresponding to the condensate densities, (d)-(f)
           Show the momentum distribution corresponding to the non-condensate 
           densities arising out of quantum fluctuations for $\Delta_x = 0, 0.4,
           1.0 a_{\rm osc}$ respectively. Here $k_x$, $k_y$ are measured 
           in units of $a_{\rm osc}^{-1}$.
          }
\label{ft0}
\end{figure}


\subsection{Momentum distribution}
\label{md}
   To relate the condensate and non-condensate density profiles with the
experimental observations based on time-of-flight imaging, it is important 
to have the momentum distribution of the atoms. For this reason, we compute
the Fourier transform of the density profiles the $k_x$-$k_y$ space. 
At $T=0$, the momentum distribution of the condensate and the
non-condensate densities have similar structures as shown in Fig.~\ref{ft0}
for different values of $\Delta_x$. For $\Delta_x=0$, the momentum 
distribution is rotationally symmetric in the $k_x$-$k_y$, but, when 
$\Delta_x\neq0$ the rotational symmetry is broken. As $\Delta_x$ is 
increased the momentum distribution along $x$- and $y$-axis broaden
and shrink, respectively. These are due to the decrease and increase in
the spatial extent of the density distributions along $x$- and $y$-axis,
respectively. Thus the change in the geometry of the trapping potential from 
multiply to simply connected is reflected in the momentum distribution. 

\begin{figure}[t]
 \includegraphics[width=8.5cm]{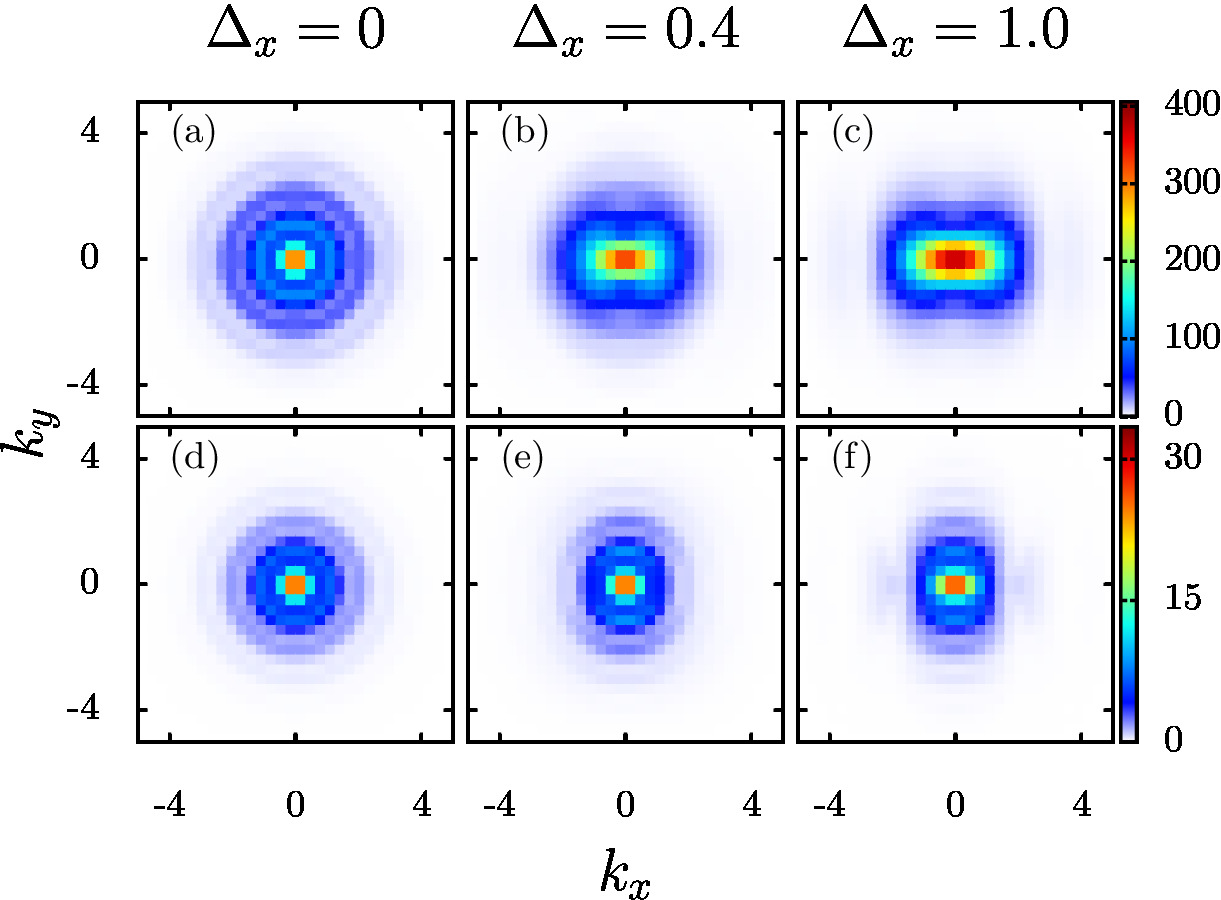}
 \caption {Momentum distribution of the densities for different values
           of $\Delta_x$ at $T=10$nK. (a)-(c) Show the momentum distribution
           corresponding to the condensate densities, (d)-(f)
           Show the momentum distribution corresponding to the non-condensate
           densities arising due to thermal atoms for $\Delta_x = 0, 0.4,
           1.0 a_{\rm osc}$ respectively. Here $k_x$, $k_y$ are measured  
           in units of $a_{\rm osc}^{-1}$.
          }
\label{ft10}
\end{figure}

 In the case of $T\neq0$, there is a marked difference between the momentum 
distributions of $n_c$ and $\tilde{n}$. The momentum 
distribution of $n_c$ at $T=0$ and $T\neq0$ are similar
as evident from Fig.~\ref{ft10}(a)-(c) for $\Delta_x\geqslant0$. The thermal
density $\tilde{n}$, on the other hand, has a very different momentum 
distribution compared to the quantum fluctuations. This is evident from the 
profile in $k_x$-$k_y$ space shown in Fig.~\ref{ft10}(d)-(f), and this
emanates from the difference in the profile of $\tilde{n}$; the contribution
from the thermal fluctuations is multiply connected, whereas it is
simply connected for quantum fluctuations. This difference in $\tilde{n}$
was discussed earlier, and evident from the profiles in Fig. \ref{qt}.


\subsection{Dispersion Curves}
\label{dsp}
\begin{figure}[t]
 \includegraphics[width=8.5cm]{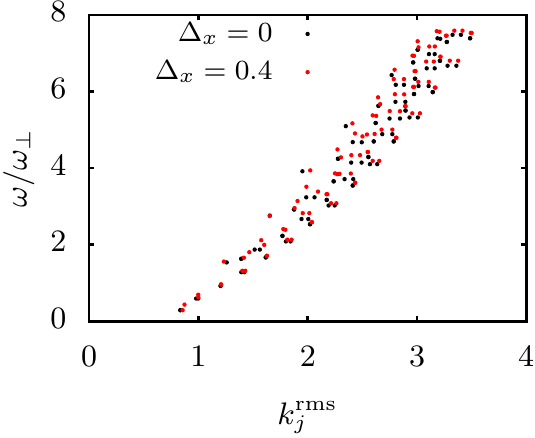}
 \caption{The BdG quasiparticle dispersion curve for the transformation
          from multiply to symmetry-breaking simply connected topology.
          Here $k^{\rm rms}_j$ is in units of $a_{\rm osc}^{-1}$.}
\label{dsprsn}
\end{figure}

 To obtain the dispersion curves using Eq.~(\ref{dspeq}) we compute
$k^{\rm rms}$ of the $j$th quasiparticle mode for different $\Delta_x$.
For $\Delta_x=0$, the azimuthal quantum number $m$ is a good quantum number,
and the modes with same $m$ form branch like structures in the dispersion 
curves. The structure is discernible in the discrete dispersion curve in 
Fig. \ref{dsprsn}. However, for $\Delta_x > 0$ there is no discernible
structure, and it is consistent with the absence of rotational symmetry in
the system. More important, with the transition from multiply to simply
connected geometry as $\Delta_x$ is changed from $0.0$ to $0.4$,
for majority of the modes $k_j^{\rm rms}$ remain unchanged, but 
the mode energies show an increase. This is a reflection of the mode hardening
of the $m\neq0$ modes discussed earlier. For a few, $m=0$ 
and $n\neq 0$ or the radially excited modes in particular, the mode energies 
do not show any significant changes, but there is an increase in 
$k_j^{\rm rms}$. The reason for this is the less number of low-energy modes 
with $n\neq0$, so a change in $u_j(\mathbf{k})$ and $v_j(\mathbf{k})$ 
due to increase in $\Delta_x$ has prominent affects on $k_j^{\rm rms}$.


\section{Conclusions}
\label{conc}
  The present studies reveal dramatic modification of the condensate, and 
thermal density profiles when the harmonic and Gaussian confining potentials 
in a toroidal trap configuration have non-coincident centers. More importantly,
larger separation of the trap centers transforms the topology of the system.
Starting from a multiply connected density profile, as the separation between 
the trap centers is increased, above a critical value a simply connected 
condensate profile emerges as the ground state configuration. An
important observation associated with this transition is the lifting of the 
degeneracy of the low-lying quasi particles, and subsequent increase in 
the mode energies. This is due to the modification of the radial trapping 
frequencies of the effective external confining potential. 

  Our finite temperature results using HFB-Popov approximation demonstrate
a contrasting trends in the quantum and thermal fluctuations as a function
of the separation of trap centers. The quantum fluctuations resembles the
condensate density distribution, and undergoes a transformation from multiply
connected to simply connected geometry. Thermal fluctuations, on the other
hand, remains multiply connected. This leads to discernible differences
in the TOF density evolutions, which could be detected in experiments.


\begin{acknowledgments}
We thank K. Suthar, S. Bandyopadhyay and R. Bai for useful
discussions. The results presented in the paper are based on the
computations using Vikram-100, the 100TFLOP HPC Cluster at Physical Research 
Laboratory, Ahmedabad, India.
\end{acknowledgments}

\bibliography{/home/arko/Documents/tex/main_bib_entry/bec}{}
\bibliographystyle{apsrev4-1}

\end{document}